\documentclass[letterpaper,twocolumn,10pt]{article}
\usepackage{usenix2019_v3_hotos}
	
\usepackage[T1]{fontenc}
\usepackage{libertine} 
\usepackage{inconsolata} 

\usepackage[utf8]{inputenc} 
\usepackage[tracking=true,activate={true,nocompatibility},final,kerning=true,spacing=true]{microtype}
\usepackage{xcolor} 
\usepackage{xspace} 
\usepackage{cite} 
\usepackage{paralist} 
\usepackage{graphicx} 
\usepackage{tikz} 
\usepackage{subcaption} 
\usepackage{tabu} 
\usepackage{makecell} 
\usepackage[usestackEOL]{stackengine} 
\usepackage{enumitem} 
\usepackage{pifont} 
\usepackage{stmaryrd} 
\usepackage{relsize} 
\usepackage[most]{tcolorbox} 
\tcbuselibrary{breakable}
\usepackage{adjustbox} 
\usepackage{array} 
\usepackage{multirow} 
\usepackage{verbatimbox} 
\usepackage{caption} 
\usepackage{layouts} 
\usepackage{bm} 
\usepackage{epigraph} 
\usepackage[hang,flushmargin,bottom]{footmisc} 

\usepackage{quoting}
\quotingsetup{vskip=0pt}

\usetikzlibrary{trees} 
\usetikzlibrary{arrows,arrows.meta,bending,intersections,decorations.markings,shapes.geometric,decorations.pathreplacing,calc}
\usetikzlibrary{fit,patterns,plotmarks,backgrounds}
\usetikzlibrary{positioning}

\makeatletter
\g@addto@macro\normalsize{%
  \setlength\abovedisplayskip{0pt}
  \setlength\belowdisplayskip{0pt}
  \setlength\abovedisplayshortskip{0pt}
  \setlength\belowdisplayshortskip{0pt}
}

\usepackage{url}
\usepackage[pdftex,colorlinks]{hyperref}
\usepackage[hyperref]{backref}
\usepackage[nameinlink,noabbrev]{cleveref} 
\DeclareCaptionLabelSeparator{forcespace}{~} 

\definecolor{figurecolor}{RGB}{22,90,220}
\definecolor{citecolor}{RGB}{198,81,19}
\renewcommand*{\backrefalt}[4]{\ifcase #1 \or (\S #2). \else (\S #2). \fi}

\usepackage{url}
\makeatletter
\g@addto@macro{\UrlBreaks}{\UrlOrds}
\makeatother

\def\Snospace~{\S{}}

\usepackage{array}
\newcolumntype{L}[1]{>{\raggedright\let\newline\\\arraybackslash\hspace{0pt}}m{#1}}
\newcolumntype{C}[1]{>{\centering\let\newline\\\arraybackslash\hspace{0pt}}m{#1}}
\newcolumntype{R}[1]{>{\raggedleft\let\newline\\\arraybackslash\hspace{0pt}}m{#1}}

\usepackage[compact]{titlesec}

\titlespacing*{\section}{0pt}{2ex}{1ex}
\titlespacing*{\subsection}{0pt}{1.4ex}{0.9ex}
\titlespacing*{\subsubsection}{0pt}{1.2ex}{0.8ex}

\hypersetup{ colorlinks=true, urlcolor=black, linkcolor=figurecolor, citecolor=citecolor, pdfstartview=FitH,
        pdftitle={No DNN Left Behind: Improving Inference in the Cloud with Multi-Tenancy},
        pdfauthor={Amit Samanta, Suhas Shrinivasan, Antoine Kaufmann, Jonathan Mace}
        }

\author{{\rm Amit Samanta, Suhas Shrinivasan, Antoine Kaufmann, Jonathan Mace}\\[1mm]{\rm Max Planck Institute for Software Systems}}
\date{}

\begin{document}

\title{No DNN Left Behind: Improving Inference in the Cloud with Multi-Tenancy}

\maketitle

\newcommand{\eg}{\emph{e.g.}\xspace}
\def\etc{etc.\@\xspace}
\newcommand{\cf}{{cf.}\xspace}
\newcommand{\ie}{\emph{i.e.}\xspace}
\newcommand{\etal}{\emph{et al.}\xspace}

\newcommand{\code}[1]{\texttt{#1}}
\newcommand{\java}[1]{\scalebox{1.0}[1.0]{{\textls[-75]{\texttt{#1}}\xspace}}}

\newcommand{\prule}{\vspace{2mm}\hrule\vspace{2mm}}  

\newcommand{\fakepara}[1]{\vspace{1.5mm}\noindent\textbf{#1}}

\newcommand{\colorcirc}[2]{\protect\raisebox{-0.5pt}{\tikzexternaldisable\protect\tikz{\protect\path [fill=#1] (0,0.004) circle [radius=0.105];\protect\node [inner sep=0,outer sep=0] at (0,0) {\small{#2}};}}}

\newcommand\smalldots{{\relsize{-1}\hspace{-0.1em}.\hspace{-0.2em}.\hspace{-0.2em}.\hspace{-0.12em}}}
\newcommand\tinydots{\textls[-30]{\relsize{-1}\hspace{-0.15em}.\hspace{-0.3em}.\hspace{-0.3em}.\hspace{-0.175em}}}

\SetTracking[spacing={-50*,0*,50*}]{encoding=T1, family=zi4}{-40}
\renewcommand\UrlFont{\fontfamily{zi4}\lsstyle}
\Urlmuskip=-1mu\relax

\SetTracking{encoding={T1}, shape=sc}{-10}

\setlength{\skip\footins}{1mm} 
\setlength{\footnotesep}{1mm} 

\begin{abstract}

With the rise of machine learning, inference on deep neural networks (DNNs) has
become a core building block on the critical path for many cloud applications.
Applications today rely on isolated ad-hoc deployments that force users to
compromise on consistent latency, elasticity, or cost-efficiency,
depending on workload characteristics.
We propose to elevate DNN inference to be a first class cloud primitive provided
by a shared multi-tenant system, akin to cloud storage, and cloud databases.
A shared system enables cost-efficient operation with consistent performance
across the full spectrum of workloads.
We argue that DNN inference is an ideal candidate for a multi-tenant system
because of its narrow and well-defined interface and predictable
resource requirements.

\end{abstract}

\section{Motivation}
\label{sec:motivation}

Deep neural networks (DNNs) excel at a wide range of machine learning tasks including computer vision, natural language processing, speech detection, and more.  
The success of DNNs has correspondingly led to the rapid growth of systems and platforms for deep learning (DL).
Today, a rich ecosystem of platforms, libraries, and runtimes make it easy to develop, train, and deploy DNNs.

In a cloud and datacenter setting, machine learning workloads have thus grown in prominence.
Broadly speaking, we can divide DL workloads into \emph{training} workloads and \emph{inference} workloads. Training is a compute-intensive batch task that constructs a DNN using large quantities of data; training bears similarity to other batch tasks like data analytics jobs and faces similar challenges.  In contrast, \emph{inference} is a low-latency, online task that generates predictions on-demand using a trained DNN; inference bears similarity to online applications like databases, web services, and microservices, and is often just one piece of a broader end-to-end application.
DNNs are typically hosted separately from application logic and accessed via remote procedure call (RPC).

In this work we consider inference workloads exclusively.  Although inference has not gone unnoticed in prior work,
existing cloud infrastructure for inference workloads still has several limitations.
These limitations are not fundamental, and we believe there are significant opportunities to better serve inference workloads.
Throughout this paper we focus on deep neural networks (DNNs) as they are the main driver of today's machine learning trends.  However, many of our observations generalize beyond DNNs.

\subsection{A Brief Primer on DNNs}
\label{sec:primer}

For our purposes it suffices to think of a DNN as a stateless, deterministic, black-box function.  For example resnet18, a DNN for image classification, takes as input a $224 \times 224 \times 3$ RGB image, performs \textasciitilde2 billion flops, and outputs a 1000-dimension vector of class probabilities~\cite{he2016deep}.

Internally, DNNs are straightforward, comprising a sequence of statically-defined \emph{layers}.  Each layer is a mathematical function that transforms the outputs from the previous layer to produce inputs for the next layer.  For example, a \emph{fully-connected layer} multiplies the output tensor of the previous layer with a tensor of hard-coded values (these were `learned' during training).  DNN depths vary from a few layers to a few hundred layers, and typically draw from a catalogue of several dozen layer types.  

Evaluating a DNN thereby entails performing the operation of each layer in turn, transforming the original input into the output.  DNNs can be evaluated on CPUs, GPUs, or special-purpose accelerators like TPUs~\cite{jouppi2017datacenter}.  GPUs and TPUs see significant speedups from parallelism; for example, for resnet18 with TVM~\cite{chen2018tvm} we measure 190.80ms median inference latency on a single-core CPU, compared to 0.97ms on a Tesla V100 GPU.

\subsection{Deploying a DNN for inference}
\label{sec:deploying}

The conventional approach to deploying a DNN is to provision a container or virtual machine (VM), and within the container or VM, host the model on a model server.  Model servers are analogous to webservers,
and 
notable examples include TensorFlow Serving~\cite{olston2017tensorflowserving}, Apache MXNet Model Server~\cite{lupesko2017modelserver}, and Clipper~\cite{crankshaw2017clipper}.  

Aside from dedicated VMs,\footnote{
In the rest of the paper we use VMs to refer to both containers and VMs for brevity.}
cloud customers can alternatively use a hosted system to serve their model, such as Google ML Engine~\cite{mlengine}
and Microsoft Azure ML~\cite{azureml}.  These systems expose a higher-level interface for users -- upload your model, and receive an endpoint to which you can send inference requests. 
 Managed systems ease deployment complexity, as users do not have to manually provision specific resources or interact with underlying VMs.   Internally, these systems also serve models using VMs.  The main appeal of hosted solutions is to avoid capacity planning -- these systems will automatically provision additional VMs if the workload increases, and alternatively tear down VMs if they are unused.
Internal systems at companies such as Facebook, Google, and Uber take a similar approach~\cite{hazelwood2018applied,olston2017tensorflowserving,hermann2017michelangelo,hermann2018scaling}.

\textls[-20]{The common strand among these approaches is isolation at the VM level.  Some model servers support deploying more than one model at a time, but this is statically configured on start-up, with pre-allocated resources~\cite{olston2017tensorflowserving}. 
All frameworks support models with custom user-code layers.}

\subsection{Expectations vs. Reality}
\label{sec:reality}

All existing approaches to deploying DNNs use VMs for isolation.  However, we argue that VMs are inefficient, undesirable, and fundamentally mismatched with the expectations and requirements of inference workloads.  Inference workloads are online workloads and are often part of broader latency-sensitive end-to-end applications.
The exact number of inference requests per second will vary unpredictably over time, often at fine time scales; meanwhile, the workload may have tail latency targets on the order of milliseconds, such as 7ms at Google~\cite{jouppi2017datacenter}, 10ms at Uber~\cite{hermann2017michelangelo}, and 25ms at Zendesk~\cite{cheeyau2017zendesk}.

\fakepara{Reality: idle resources and over-provisioning}
With this in mind, the basic approach of statically provisioning VMs has clear drawbacks.  First and foremost, users must either (i) over-provision to satisfy the estimated peak demand, thus leaving resources idle much of the time; or (ii) accept increased latency and even denial of service if demand increases.  Moreover, some workloads may \emph{never} have sufficient demand to warrant an entire VM, and may even undergo long periods of idleness.  Statically-provisioned VMs are wasteful, as users must nonetheless provision and pay for the excess capacity.  The problem is further compounded when using expensive hardware accelerators like GPUs and TPUs, which only make sense for workloads that can sustain thousands of requests per second~\cite{jouppi2017datacenter}.
Lastly, mapping workload requirements to concrete resources is non-trivial.  In a recent Amazon blog post, authors describe how \emph{``developers are often stumped when the time comes to pick an instance type and size. Indeed, for larger models, the inference latency of CPUs may not meet the needs of online applications, while the cost of a full-fledged GPU may not be justified.''}~\cite{simon2018elasticinference}

\fakepara{Reality: scaling in coarse-grained increments}
\textls[-17]{Statically-provisioned VMs are further mismatched for workloads with tight latency requirements.  To satisfy these latency requirements, it may be necessary to use hardware accelerators like GPUs and TPUs.  Hardware accelerators can provide several orders of magnitude speedup for inference workloads; for example, for resnet18 we measure 190.80ms median inference latency on one-core Google Cloud VM, compared to 0.97ms on a Tesla v100.  However, hardware acceleration is both more expensive and more \emph{coarse-grained} than CPU-only VMs.  Concretely, the throughput of resnet18 on one CPU core is 5.24 inferences per second (inf/s), at a cost of 3.48c/hr (\$1.84 per million inferences).  Conversely the throughput of resnet18 on a GPU is 1031 inf/s.\footnote{With batching and batch-interleaving we can increase throughput up to 4083 inf/s at the cost of elevated 62.1ms median latency.}  
GPUs are therefore more cost-effective at \$2.55/hr (\$0.69 per million inferences).
However, there is clearly a tradeoff: though GPUs have the potential for better latency and throughput, CPUs are scalable in much finer increments.  Each generation of GPU and CPU offers a different point on this scale.}

\fakepara{Reality: slow auto-scaling}
Beyond static resources, cloud providers also offer managed solutions, such as Google ML Engine, and Microsoft Azure ML, outlined in~\autoref{sec:deploying}.  These solutions will scale provisioned resources in response to fluctuations in the workload.  To do so, customers must specify latency targets or throughput thresholds, which trigger the system to automatically spin up new VMs or tear down VMs as workloads change.  However, VMs are inherently slow to spin up, and as such, auto-scaling can only adapt to workload fluctuations over long time periods.  For example, Amazon SageMaker~\cite{sagemaker} makes scaling decisions in 5-minute increments by default.  Google Cloud ML's documentation reflects this limitation: ``\emph{If your traffic regularly has steep spikes, and if reliably low latency is important to your application, you may want to consider manual scaling.}''~\cite{mlengine}.

\fakepara{Reality: high cold-start latency}
As a complement to auto-scaling, managed solutions also enable VMs to be torn down completely if workloads undergo long periods of idleness.  This avoids paying for resources that will not be used, albeit over long time scales.  However, when a workload picks up again, the resources must be re-provisioned.  This task, sometimes referred to as \emph{cold-start}, can take many seconds.  For example, researchers evaluating the feasibility of DNNs in serverless applications measured cold-start times of up to 12 seconds for 100MB models~\cite{ishakian2018serving}.  This increased latency is clearly at odds with the needs of online workloads, and in particular, any workload that may be infrequent or sporadic.

\fakepara{Reality: mismatched pricing abstractions}
Although managed solutions offer a high-level abstraction for inference workloads -- ``give us your model, then send us your inference requests'' -- this abstraction is not reflected in the pricing models of these systems.  Both the underlying implementations of these managed solutions, and subsequent pricing models, are based on total VM time, \emph{including idle time}, instead of per inference or total compute time, as is common in, \eg serverless computing environments~\cite{azurefunctions}.  Consequently, the costs of idle resources due to over-provisioning are reflected back on the user.  Given the long time scales over which these systems make decisions -- \eg waiting 10 minutes before tearing down a Cloud ML container~\cite{cloudmlcosts} -- this adds significant financial cost to users.  This represents a fundamental tradeoff that is inherent to using VMs as the unit of provisioned resource: paying for unused, idle resources vs. incurring high cold-start latency.

\subsection{Summary of Requirements}

Based on these challenges, we summarize the following desirable properties for inference workloads:

\fakepara{Latency} Inference workloads need stable average and tail latency, regardless of workload volume.  Infrequent or sporadic workloads should not suffer from high cold-start latency.

\fakepara{Elasticity} Online workloads are inherently unpredictable, and may rapidly transition from low-volume to high-volume.  Workload fluctuations should be handled transparently.

\fakepara{Cost-Effective}  Users should not have to significantly over-provision resources.  Moreover, costs should be consistent regardless of the workload volume, \ie users should be charged based on the work done.


\section{Multi-Tenant Systems}
\label{sec:multitenant}

VMs and containers are a poor fit for inference workloads.  Our proposed alternative is to share the underlying resources across tenants, by executing inference workloads within a \emph{shared, multi-tenant system}.  In shared systems, the system operator provisions resources for the system as a whole, and runs long-lived system processes that receive and execute requests from different tenants concurrently.  By sharing resources, fluctuations in demand can be amortized across tenants, and we avoid over-provisioning and wasting resources.  By sharing processes, workload spikes can be absorbed by re-distributing load, and workloads with long periods of inactivity do not incur cold-start latency.  And lastly, since cloud providers maintain control over the system and its resources, they can more closely align the pricing and system abstractions (\eg, by charging on a per-request basis).  Today, multi-tenant systems already provide a variety of core datacenter services, such as storage~\cite{ghemawat2003gfs,calder2011windows}, databases~\cite{hbase,chang2006bigtable}, queueing~\cite{kafka}, and co-ordination~\cite{burrows2006chubby,hunt2010zookeeper}.

Multi-tenant systems are only justifiable for \emph{core datacenter functionality}, where there is a common need for the functionality across many tenants and workloads.
We believe that DNN inference is sufficiently important and prevalent to justify a specialized multi-tenant system.  A shared system can significantly improve elasticity and cost-effectiveness.  However, in designing a multi-tenant system for inference workloads, we must juggle our original goals with three additional challenges:

\fakepara{Abstractions} Tenants no longer interact with compute resources directly.  Instead, shared systems expose high-level abstractions and interfaces.  These abstractions \emph{must} generalize across many workloads and tenants.

\fakepara{Security} Shared systems execute requests of different tenants within the same, shared processes.  Thus, users are no longer separated by rigid OS or hypervisor boundaries.  Nonetheless, we must still ensure security between different tenant workloads.

\fakepara{Performance Isolation} The system must prevent performance interference between different tenants.
Performance isolation is arguably the most difficult challenge for multi-tenant systems today, and remains an active area of research for multi-tenant systems in general~\cite{mace2014towards,mace2015retro,mace20162dfq,yang2018principled, angel2014end}.

\vspace{2mm}
In the remainder of this paper, we outline how a multi-tenant system for DNN inference can address these challenges, while also satisfying our original motivating requirements.

\section{Design of a Shared Inference System}
\label{sec:design}

Our proposed system architecture bears similarity to many other shared
systems
~\cite{ghemawat2003gfs,chang2006bigtable,hbase}.  High-level meta-operations are handled by a logically centralized controller.  Meanwhile, DNN hosting and inference are handled by worker processes, spread across many machines.

\textls[-10]{Users first upload a DNN to the system and then start sending inference requests.
Internally, the system distributes the DNN to one or more workers.
Then, inference requests are routed to the workers that host the DNN.
Workers host DNNs for many tenants simultaneously, and receive incoming requests.
If a DNN is in high-demand, it is replicated across many workers.
Replication achieves both fault tolerance, and elasticity.
In our design, we focus particular attention on how workers multiplex inference requests across potentially many models.}

\subsection{Inference Runtime}

Since the focus of our system is inference, a pre-requisite is for users to have trained their DNN using an existing deep learning framework such as TensorFlow~\cite{abadi2016tensorflow}.  
However, today's DL frameworks are designed to handle the entire model lifecycle, from training to deployment, and crucially, the framework also provides the runtime for model execution.
In a multi-tenant system it is infeasible to allow users to upload an arbitrary runtime for executing their model, as this would require heavyweight isolation techniques to guarantee security, and it would limit opportunities for optimization.  Instead, our system contains an internal model execution runtime and does not require user-code for performing inference.

However, using a custom runtime means users must submit models in a format understood by the runtime.
Recent efforts in the machine learning community to address framework inter-operability have led to the development of the Open Neural Network eXchange (ONNX) model format~\cite{onnx}.
The ONNX model format is simply a high-level description of the structure and parameters of a trained DNN, without framework-specific code or runtime optimizations.

The downside of this abstraction is that it restricts users to a pre-defined set of supported layer types.  While existing DL frameworks allow users to implement custom layer types, user extensions are not feasible in a multi-tenant system.
We do not believe that this is a significant limitation for our system targeting common-case workloads.
For example, 95\% of production DNNs at Google use standard layers; moreover, Google's production hardware accelerator, the TPU, only supports standard layers~\cite{jouppi2017datacenter}.
Thus, supporting prototype layers is unnecessary until they reach mainstream adoption, at which point they can be included in our set of supported layer types.
In general this restriction is reasonable and multi-tenant systems often disallow custom user code; for example, multi-tenant databases typically do not support SQL's user-defined functions~\cite{hbase, calder2011windows}.

\subsection{Performance Isolation}
\label{sec:perfisolation}

In shared systems, performance isolation is important for ensuring aggressive tenants or unpredictable workloads do not cause starvation, reduced throughput, or high latency for others.  However, comprehensive performance isolation is a challenge, even for existing systems today. 
Difficulties arise because 
isolation must be implemented at the application level, where we lack the ability to pre-empt requests while they are executing.  
A common approach is to predict resource requirements, measure actual consumption, and use coarse-grained feedback loops to provide guarantees like fairness over time~\cite{mace2015retro, shue2012performance}.  Often this is implemented as a fair queue scheduler at the request admission point~\cite{mace20162dfq}.

However, unlike workloads from other domains, DNN inference has highly predictable resource consumption patterns.  In other multi-tenant systems, performance isolation is difficult primarily because resource requirements are unpredictable and vary widely from request to request~\cite{mace20162dfq}, and once a request is admitted it runs to completion.  DNN inference does not face this challenge, because inference is a fundamentally predictable computation.  This stems from the structure of DNNs (\autoref{sec:primer}) -- they are a fixed sequence of mathematical operations.  A priori, we can quantify the exact number of flops required by each layer of the DNN.  Moreover, DNNs are predictable as they do not contain control flow elements.\footnote{This does not preclude higher-level control flow, which is the subject of recent research~\cite{yu2018dynamic, jeong2018improving}}  DNNs that accept variable-sized or batched inputs also vary deterministically based on input size.

We can exploit this predictability to do a much better job of scheduling, whether at request admission, or at finer granularity within the system.  Instead of modeling costs up front, we propose a more pragmatic approach based on measurement.  In our experiments with TVM~\cite{chen2018tvm} we measure 99th percentile latencies not exceeding 15\% of the mean for a range of off-the-shelf DNNs~\cite{onnxmodelzoo} and workload mixes.

Predictable computations enable systems to react to workload fluctuations much more quickly, and enable higher quality scheduling decisions.  For example, instead of heuristic-based best effort scheduling, an admission scheduler can confidently optimize an objective across all pending requests, such as minimizing average latency.  Overall, predictable DNN inference presents an opportunity both to improve upon existing resource management techniques, and to explore new approaches entirely.

\subsection{Efficiency and Optimizations}

A key characteristic of a multi-tenant system is to alternate service between different tenants.  In the worst case, each request may require loading and executing a different model that is not currently loaded.
This introduces additional resource costs, such as the need to copy the model from a remote machine or from cold storage.  Similarly, if we use hardware accelerators, then models need to be copied from host memory to device memory.
Overall, the total inference latency will depend on a combination of execution latency (CPU, GPU, or other accelerator) and transfer latency (PCIe, disk, and/or network).  As mentioned in~\autoref{sec:perfisolation}, execution latency is predictable; but so too is transfer latency, since the memory footprint of a DNN is fixed.  For example, resnet18 allocates approximately 78MB device memory for DNN weights; we measure \textasciitilde7ms increased latency when copying weights prior to each inference, consistent with 12GB/s PCIe bandwidth.

Not all inference requests incur memory transfer overheads because of caching opportunities at each level.  The typical memory footprint of a DNN is in the tens or hundreds of MBs; in contrast, servers often exceed 1TB of main memory, current-generation GPUs have up to 32GB device memory, and current-generation TPUs have 64GB device memory.  Most requests can exploit cached models instead of reloading from scratch.  Consequently, bottlenecks will vary based on resource requirements of each DNN and cache hit ratios.  For example, for models like resnet18, a GPU cache hit ratio of 85\% or greater would shift the bottleneck resource from PCIe bandwidth to GPU execution latency.

In all, this leads to a multi-resource scheduling problem that, while similar to work in other domains~\cite{ghodsi2012multi, mace2015retro}, has some unique constraints: (i) inference requests have predictable resource requirements; (ii) inference requests consume resources one-at-a-time; (iii) resources are independent and asynchronous; (iv) resources have measurable concurrency and throughput; and (v) scheduling decisions can be interposed before each resource.  These constraints present an opportunity for high quality, fine-grained scheduling.

Beyond fine-grained scheduling decisions, we also have opportunities for high-quality placement and load-management decisions.  For any request, we can calculate with high confidence the latency of local execution including any memory transfers.  We can consider alternative execution strategies, such as CPU execution vs. hardware accelerator, and local vs. remote.  A worker with several pending requests can calculate a priori the expected completion time of each request, including queueing time, and pre-emptively cancel or re-route requests accordingly.

Of course, the optimizations we have described primarily affect models with infrequent or varying workload patterns, for which multi-tenancy `reclaims' resources that would otherwise go unused.  This does not, however, come at the expense of degraded performance for heavy workloads.  If a model does have a heavy workload (\ie, enough demand to saturate a worker entirely), then we migrate colocated models elsewhere, giving the heavy workload essentially exclusive use of the worker.  Then, any hardware~\cite{jouppi2017datacenter} or software~\cite{gao2018low,lee2018pretzel} optimizations are equally applicable.


\section{Discussion}
\label{sec:discussion}

Multi-tenancy has complementary goals to much of the prior work around DNN inference.  Assuming sufficient workload demand from individual models, multi-tenant systems can equally benefit from specially designed accelerators~\cite{jouppi2017datacenter}, fine-grained batching techniques~\cite{gao2018low}, and potential future results in inter-model batching~\cite{lee2018pretzel}.  A multi-tenant system would be particularly well-placed for exploiting inter-model optimizations, as the system controls model placement and co-location decisions.

In the research literature, the most similar system to what we propose is Clipper~\cite{crankshaw2017clipper}.  Clipper also proposes an abstraction to serve as a ``narrow waist'' for model deployment, albeit different to our proposed abstraction, and isolates models using containers.   Clipper focuses on challenges and optimizations that lie \emph{above} their proposed interface -- model management, latency-throughput tradeoffs, and higher-level concerns like prediction accuracy.  By contrast, we consider different challenges and optimizations that lie \emph{below} our proposed interface.
In industry, the most similar system to what we propose is TFS$^2$~\cite{olston2017tensorflowserving}, Google's internal model hosting system, which distributes models to shared worker processes and provides automatic scaling; however, there is insufficient public information for a detailed comparison.

In this paper we did not discuss pre- and post-processing of DNN inputs and outputs, an important step for every DNN pipeline.  We believe that this step is better handled by a separate (but possibly co-located and co-designed) system, that composes much like \eg distributed file systems and databases.  Processing steps have different performance characteristics compared to DNN inference, and often rely on user code; of course, this does not preclude entirely the possibility of safe high-level abstractions.  The biggest difference between processing and inference is inter-model commonality.  Pre- and post-processing steps are often similar between DNNs, and can be batched, even across different model pipelines~\cite{lee2018pretzel}.  However, DNN inference has fewer opportunities for batching across models, as model weights are unique.

Lastly, DNN inference does not cover all machine learning workloads.  But, by restricting our design to this specific but common workload class, it enables assumptions around performance, predictability, and clarity, that we would otherwise lack.   Models beyond DNNs have fundamentally different performance characteristics, thus we omit them from consideration.  Similarly, we exclude reinforcement learning, which does not have a distinct inference phase.  Multi-tenant systems for these other scenarios may also make sense, and we look forward to seeing future research in this direction.

\section{Conclusion}

In this paper we proposed that DNN inference should be a first-class cloud primitive, provided by a shared multi-tenant system.  Multi-tenancy enables cost-efficient operation with consistent performance across a wide range of workloads.  DNN inference is ideally suited for multi-tenancy because of its predictable resource requirements, which help address the important question of performance isolation.

\bibliographystyle{acm}
\interlinepenalty=10000
\bibliography{paper}

\end{document}